\def\be{\begin{equation}}
\def\ee{\end{equation}}
\def\ba{\begin{array}}
\def\ea{\end{array}}

\documentclass[aps,amsmath,amssymb,amsfonts,showpacs]{revtex4}
\usepackage{graphicx}
\usepackage{amsmath,amsthm}
\def\qed{\leavevmode\unskip\penalty9999 \hbox{}\nobreak\hfill
     \quad\hbox{\leavevmode  \hbox to.77778em{%
               \hfil\vrule   \vbox to.675em%
               {\hrule width.6em\vfil\hrule}\vrule\hfil}}
     \par\vskip3pt}

\begin{document}
\title{Separability criteria via sets of mutually unbiased measurements}
\author{Lu Liu$^{1}$}
\author{Ting Gao$^{1}$}
\email{gaoting@hebtu.edu.cn}
\author{Fengli Yan$^{2}$}
\email{flyan@hebtu.edu.cn}

\affiliation{$^{1}$College of Mathematics and Information Science, Hebei
Normal University, Shijiazhuang 050024, China, $^{2}$College of Physics Science and Information Engineering, Hebei
Normal University, Shijiazhuang 050024, China}

\begin{abstract}
Mutually unbiased measurements (MUMs) are generalized from the concept of mutually unbiased bases (MUBs) and
include the complete set of MUBs as a special case, but they are superior to MUBs as they do not need to be rank one projectors.
We investigate entanglement detection using sets of MUMs and derive separability criteria for multipartite qudit systems,  arbitrary high-dimensional bipartite systems of a $d_1$-dimensional subsystem and a $d_2$-dimensional subsystem, and multipartite systems of multi-level subsystems. These criteria are of the advantages of more effective and wider application range than previous criteria.  They provide experimental implementation in detecting entanglement of unknown quantum states.
\end{abstract}


\maketitle

\vspace{0.5cm} \noindent{\bf\large Introduction }

\noindent
Quantum entanglement as a new physical resource has drawn a lot of attention in the field of quantum information in the past decade \cite{PRL70.1895, PRL67.661,PRL68.557,PR448.1,Science283.2050,PJB41.75,JPA38.5761,PRA83.022319,GaoEPL84,Nature404.247}.
It plays a significant role in quantum information processing and has wide applications such as quantum cryptography \cite{PRL67.661,77.2816,56.1163}, quantum teleportation \cite{PRL70.1895,GaoEPL84,PLA276.8,272.32,PRA66.012301,GaoCTP2004}, and dense coding \cite{PRL69.2881}.
A main task of the theory of quantum entanglement is to distinguish between entangled states and separable states.
For bipartite systems, various separability criteria have been proposed such as positive partial transposition criterion \cite{PRL77.1413}, computable cross norm or realignment criterion \cite{QIC3.193}, reduction criterion \cite{PRA59.4206}, and covariance matrix criterion \cite{PRL99.130504}.
For multipartite and high dimensional systems, this problem is more complicated. There are various kinds of classification for multipartite entanglement.
For instance, one can discuss it with the notions of $k$-partite entanglement or $k$-nonseparability for given partition and unfixed partition, respectively.
In \cite{PRA82.062113}, Gao \textit{et al} obtained separability criteria which can detect genuinely entangled and nonseparability $n$-partite mixed quantum states in arbitrary dimensional systems, and further developed $k$-separability criteria for mixed multipartite quantum states \cite{EPL.104.20007}. In \cite{PRA.86.062323}, the authors defined $k$-ME concurrence in terms of all possible $k$ partitions, which is a quantitative entanglement measure that has some important properties. One of the most important property is that $C_{k-\mathrm{ME}}$ is zero if and only if the state is $k$ separable. Combining $k$-ME concurrence with permutation invariance, a lower bound was given on entanglement for the permutation-invariance part of a state that apply to arbitrary multipartite states \cite{PRL.112.180501}. At the same time, the concept of ``the permutationally invariant (PI) part of a density matrix" is proven to be more powerful because of its basis-dependent property.

Although there have been numerous mathematical tools for detecting entanglement of a given known quantum state, fewer results were obtained of the experimental implementation of entanglement detection for unknown quantum states. In 1960, Schwinger introduced the notion of mutually unbiased bases (MUBs) under a different name \cite{1409.3386[7]}. He noted that mutually unbiased bases represent maximally non-commutative measurements, which means the state of a system described in one mutually unbiased base provided no information about the state in another.

Later the term of mutually unbiased bases were introduced in \cite{1409.3386[11]}, as they are intimately related to the nature of quantum information \cite{PRL.110.143601[3],PRL.110.143601[4],PRL.110.143601[5]}.
Entanglement detection using entropic uncertainty relations for two MUBs was developed in \cite{NP.6.659} and extended to arbitrary numbers of MUBs in \cite{PRA.90.062127}. This method was experimentally implemented in \cite{NP.7.757}.
In \cite{PRA.86.022311}, the authors availed of mutually unbiased bases and obtained separability criteria in two-qudit, multipartite and continuous-variable quantum systems.
For two $d$-dimensional systems, the criterion is shown to be both necessary and sufficient for the separability of isotropic states when $d$ is a prime power. However, when $d$ is not a prime power, the criterion becomes less effective.
The maximum number $N(d)$ of mutually unbiased bases has been shown to be $d+1$ when $d$ is a prime power, but the maximal number of MUBs remains open for all other dimensions \cite{1409.3386[11]}, which limits the applications of mutually unbiased bases.
The concept of mutually unbiased bases were generalized to mutually unbiased measurements (MUMs) in \cite{NJP.16.053038}. A complete set of $d+1$ mutually unbiased measurements were constructed \cite{NJP.16.053038} in a finite, $d$-dimensional Hilbert space, no matter whether $d$ is a prime power.
Recently, Chen, Ma and Fei connected the separability criteria to mutually unbiased measurements \cite{PRA.89.064302} for arbitrary $d$-dimensional bipartite systems. Another method of entanglement detection in bipartite finite dimensional systems were realized using incomplete sets of mutually unbiased measurements \cite{1407.7333}. In \cite{1407.7333}, the author derived entropic uncertainty relations and realized a method of entanglement detection in bipartite finite-dimensional systems using two sets of incomplete mutually unbiased measurements.

In this paper, we study the separability problem via sets of mutually unbiased measurements and propose separability criteria for the separability of  multipartite qudit systems,  arbitrary high dimensional bipartite systems and multipartite systems of multi-level subsystems.

\vspace{0.5cm} \noindent{\bf\large Preliminaries }

\noindent
Two orthonormal bases $\mathcal{B}_{1}=\{|b_{1i}\rangle\}_{i=1}^{d}$ and $\mathcal{B}_{2}=\{|b_{2i}\rangle\}_{i=1}^{d}$ in Hilbert space $\mathbb{C}^{d}$ are called \emph{mutually unbiased} if and only if
\begin{equation}\label{}
|\langle b_{1i}|b_{2j}\rangle|=\frac{1}{\sqrt{d}} ,~~~\forall\, i,j=1,2,\cdots,d.
\end{equation}
A set of orthonormal bases $\{\mathcal{B}_{1},\mathcal{B}_{2},\cdots,\mathcal{B}_{m}\}$ of Hilbert space $\mathbb{C}^{d}$
 is called a set of \emph{mutually unbiased bases} (MUBs) if and only if every pair of bases in the set is mutually unbiased.
If two bases are mutually unbiased, they are maximally non-commutative, which means a measurement over one such basis leaves one completely uncertain as to the outcome of a measurement over another one, in other words, given any eigenstate of one, the eigenvalue resulting from a measurement of the other is completely undetermined. If $d$ is a prime power, then there exist $d+1$ MUBs, which is a complete set of MUBs, but the maximal number of MUBs is unknown for other dimensions. Even for the smallest non-prime-power dimension $d=6$, it is unknown whether there exists a complete set of MUBs \cite{1409.3386[11]}. For a two qudit separable state $\rho$ and any set of $m$ mutually unbiased bases $\mathcal{B}_{i}=\{|b_{ij}\rangle\}_{j=1}^{d},i=1,2,\cdots,m$, the following inequality
\begin{equation}\label{}
 I_{m}(\rho)=\sum\limits_{i=1}^{m}\sum_{j=1}^{d}\langle b_{ij}|\otimes\langle b_{ij}|\rho|b_{ij}\rangle\otimes|b_{ij}\rangle
\leq 1+\frac{m-1}{d}
\end{equation}
holds \cite{PRA.86.022311}. Particularly, for a complete set of MUBs, the inequality above can be simplified as $I_{d+1}\leq 2$.

To conquer the shortcoming that we don't know whether there exists a complete set of MUBs for all dimensions, Kalev and Gour generalized the concept of MUBs to mutually unbiased measurements (MUMs) \cite{NJP.16.053038}.
Two measurements on a $d$-dimensional Hilbert space,
${\mathcal{P}}^{(b)}=\{P^{(b)}_n|P^{(b)}_n\geq0,~\sum\limits_{n=1}^{d}P^{(b)}_n=I\}$, $b{=}1,~2$, with $d$ elements each, are said to be \emph{mutually unbiased measurements} (MUMs) \cite{NJP.16.053038} if and only if,
\begin{equation}
\begin{split}
\mathrm{Tr}(P^{(b)}_n)=&1,\\
\mathrm{Tr}(P^{(b)}_n P^{(b')}_{n'})=&\delta_{n,n'}\delta_{b,b'}\kappa
+(1-\delta_{n,n'})\delta_{b,b'}\frac{1-\kappa}{d-1}
+(1-\delta_{b,b'})\frac 1{d}.\\
\end{split}
\end{equation}
Here $\kappa$ is efficiency parameter, and $\frac{1}{d}<\kappa\leq 1$.

A complete set of $d+1$ MUMs in $d$ dimensional Hilbert space were constructed in \cite{NJP.16.053038}. Consider $d^{2}-1$  Hermitian, traceless operators acting on $\mathbb{C}^{d}$ satisfying $\mathrm{Tr}(F_{n,b}F_{n',b'})=\delta_{n,n'}\delta_{b,b'}$. Here, the generators of $SU(d)$ were used \cite{NJP.16.053038}
\begin{align}
F_{n,b}=\begin{cases}
  \frac{1}{\sqrt{2}}(|n\rangle\langle b|+|b\rangle\langle n|),& \text{for } n<b, \\
   \frac{i}{\sqrt{2}}(|n\rangle\langle b|-|b\rangle\langle n|),& \text{for } b<n, \\
   \frac1{\sqrt{n(n+1)}}(\sum\limits_{k=1}^n|k\rangle\langle k|\\
   -n|n+1\rangle\langle n+1|),& \text{for } n=b,
    \text{with }n=1,2,\cdots,d-1.
    \end{cases}
\end{align}
Using such operators,  a set of traceless, Hermitian operators $F_{n}^{(b)}$, $b=1,2,\cdots,d+1$, $n=1,2,\cdots,d,$ were built as follows \cite{NJP.16.053038},
\begin{equation}
F_{n}^{(b)}=
\begin{cases}
   F^{(b)}-(d+\sqrt{d})F_{n,b},&n=1,2,\cdots,d-1;\\[2mm]
   (1+\sqrt{d})F^{(b)},&n=d,
\end{cases}
\end{equation}
where $F^{(b)}=\sum\limits_{n=1}^{d-1}F_{n,b}$, $b=1,2,\cdots,d+1.$
Then one can construct $d+1$ MUMs explicitly \cite{NJP.16.053038},
\begin{equation}\label{3}
P_{n}^{(b)}=\frac{1}{d}I+tF_{n}^{(b)},
\end{equation}
where $t$ is chosen such that $P_{n}^{(b)}\geq0$.
These operators $\{F^{(b)}_n\}$ satisfy the conditions \cite{NJP.16.053038},
\begin{equation}\label{}
\begin{array}{ll}
& \mathrm{Tr}(F^{(b)}_n F^{(b)}_{n'})=(1+\sqrt{d})^2[\delta_{nn'}(d-1)-(1-\delta_{nn'})],\\
& \sum\limits_{n=1}^d F^{(b)}_n=0,\\
& \mathrm{Tr}(F^{(b)}_n F^{(b')}_{n'})=0,~\forall b\neq b', \;~\forall n,n'=1,2,\cdots,d.
\end{array}
\end{equation}

Given a set of $M$ MUMs $\mathbb{P}=\{\mathcal{P}^{(1)},\cdots,\mathcal{P}^{(M)}\}$ of the efficiency $\kappa$ in $d$ dimensions, consider the sum of the corresponding indices of coincidence for the measurements, there is the following bound \cite{1407.7333},
\begin{equation}
\sum\limits_{\mathcal{P}\in\mathbb{P}}C(\mathcal{P}|\rho)\leq\frac{M-1}{d}+\frac{1-\kappa+(\kappa d-1)\textrm{Tr}(\rho^{2})}{d-1},\label{M-1}
\end{equation}
where $C(\mathcal{P}^{(i)}|\rho)=\sum\limits_{n=1}^{d}[\textrm{Tr}(P_{n}^{(i)}\rho)]^{2}$, $\mathcal{P}^{(i)}=\{P_{n}^{(i)}\}_{n=1}^{d}$, $i=1,2,\cdots,M$.
For the complete set of $d+1$ MUMs, we actually have an exact result instead of the inequality \cite{1407.6816},
\begin{equation}
\sum\limits_{b=1}^{d+1}C(\mathcal{P}^{(b)}|\rho)=1+\frac{1-\kappa+(\kappa d-1)\textrm{Tr}(\rho^{2})}{d-1}.
\end{equation}
For pure state the equation can be more simplified as
\begin{equation}
\sum\limits_{b=1}^{d+1}C(\mathcal{P}^{(b)}|\rho)=1+\kappa.\label{1+k}
\end{equation}
Corresponding to the construction of MUMs, the parameter $\kappa$ is given by
\begin{equation}
\kappa=\frac{1}{d}+t^{2}(1+\sqrt{d})^{2}(d-1).
\end{equation}

\vspace{0.5cm} \noindent{\bf\large Detection of multipartite entanglement }

\noindent
For multipartite systems, the definition of separability is not unique. So we introduce the notion of $k$-separable first. A pure state $|\varphi\rangle\langle\varphi|$ of an $N$-partite is $k$-separable if the $N$ parties can be partitioned into $k$ groups
$A_{1},A_{2},\cdots,A_{k}$ such that the state can be written as a tensor product
$|\varphi\rangle\langle\varphi|=\rho_{A_{1}}\otimes\rho_{A_{2}}\otimes\cdots\otimes\rho_{A_{k}}$. A general mixed state $\rho$ is $k$-separable if it can be written as a mixture of $k$-separable states $\rho=\sum\limits_{i}p_{i}\rho_{i}$, where $\rho_{i}$ is $k$-separable pure states. States that are $N$-separable don't contain any entanglement and are called fully separable. A state is called $k$-nonseparable if it is not $k$-separable, and a state is 2-nonseparable if and only if it is genuine $N$-partite entangled. Note that the definitions above for $k$-separable mixed states don't require that each $\rho_{i}$ is $k$-separable under a fixed partition. But in this paper, we consider $k$-separable mixed states as a convex combination of $N$-partite pure states, each of which is $k$-separable with respect to a fixed partition. The notion of fully separable are same in both statements. In the following theorems, we give the necessary conditions of fully separable states. For $k$-separable state for given partition we will discuss it after the theorems.

Firstly, we will give a lemma that is generalized from the AM-GM inequality \cite{website}.

\vspace{0.2cm} \noindent\textbf{Lemma 1.}~~{\slshape
For any list of $n$ nonnegative real numbers $x_{1},x_{2},\cdots,x_{n}$, we have the following inequality
\begin{equation}
x_{1}x_{2}\cdots x_{n}\leq\Big(\frac{x_{1}^{2}+x_{2}^{2}+\cdots+x_{n}^{2}}{n}\Big)^{\frac{n}{2}}.
\end{equation}
}

\noindent\textbf{Proof.}~
Since the AM-GM inequality \cite{website}
\begin{equation}\label{}
\sqrt[^n\!]{a_{1}a_{2}\cdots a_{n}}\leq\frac{a_{1}+a_{2}+\cdots+a_{n}}{n},
\end{equation}
where $a_{1},a_{2},\cdots,a_{n}$ are any list of $n$ nonnegative real numbers,
and the equality holds if and only if $a_{1}=a_{2}=\cdots=a_{n}$.
 For $x_{1},x_{2},\cdots,x_{n}$, we have
\begin{equation}\label{}
\sqrt[\uproot{8}n]{x_{1}^{2}x_{2}^{2}\cdots x_{n}^{2}}\leq\frac{x_{1}^{2}+x_{2}^{2}+\cdots+x_{n}^{2}}{n},
\end{equation}
that is
\begin{equation}\label{}
 (x_{1}x_{2}\cdots x_{n})^{\frac{2}{n}}\leq\frac{x_{1}^{2}+x_{2}^{2}+\cdots+x_{n}^{2}}{n}.
\end{equation}
The function $f(x)=x^{a}$ is an increasing function when $a\geq0$ and $x\geq0$, so for nonnegative real numbers $x_{i},~i=1,2,\cdots,n$, we have
\begin{equation}\label{}
x_{1}x_{2}\cdots x_{n}\leq\Big(\frac{x_{1}^{2}+x_{2}^{2}+\cdots+x_{n}^{2}}{n}\Big)^{\frac{n}{2}},
\end{equation}
which completes the proof.

\vspace{0.2cm} \noindent\textbf{Theorem 1.}~~{\slshape
Let $\rho$ be a density matrix in $(\mathbb{C}^{d})^{\otimes m} $ and $\{\mathcal{P}^{(b)}_{i}\}$ be any $m$ sets of $M$ MUMs on $\mathbb{C}^{d}$ with efficiency $\kappa_{i}$,
 where $\mathcal{P}^{(b)}_{i}=\{P^{(b)}_{i,n}\}^{d}_{n=1}$.
Define $J(\rho)=\sum\limits_{b=1}^{M}\sum\limits_{n=1}^{d}
\mathrm{Tr}[(\otimes^{m}_{i=1} P_{i,n}^{(b)}) \rho].$
If $\rho$ is fully separable, then
\begin{equation}
J(\rho)\leq \frac{M-1}{d}+ \frac{1}{m}\sum\limits_{i=1}^{m}\kappa_{i}.
\end{equation}

}

\vspace{0.2cm}

\noindent\textbf{Proof.}~To prove that the inequality is satisfied for all fully separable states, let us verify that it holds for any fully separable pure state $\rho=\otimes_{i=1}^{m}|\psi_{i}\rangle\langle\psi_{i}|$ first. Note that
\begin{equation}
\begin{array}{ll}
&J(\rho)=\sum\limits_{b=1}^{M}\sum\limits_{n=1}^{d}
              \textrm{Tr}\Big[(\otimes_{i=1}^{m}P_{i,n}^{(b)})\rho\Big]\\
&~~~~~~=\sum\limits_{b=1}^{M}\sum\limits_{n=1}^{d}
   \prod_{i=1}^{m}\textrm{Tr}(P_{i,n}^{(b)}|\psi_{i}\rangle\langle\psi_{i}|)
   \end{array}
\end{equation}
and $0\leq \textrm{Tr}(P_{i,n}^{(b)}|\psi_{i}\rangle\langle\psi_{i}|)\leq 1$, by using Lemma 1, we have
\begin{equation}
\begin{array}{ll}
&J(\rho)\leq \sum\limits_{b=1}^{M}\sum\limits_{n=1}^{d}
\Big\{\frac{1}{m} \sum\limits_{i=1}^{m}\Big[\textrm{Tr}(P_{i,n}^{(b)}|\psi_{i}\rangle\langle\psi_{i}|)\Big]^2\Big\}^{\frac{m}{2}}\\
&~~~~~~\leq \sum\limits_{b=1}^{M}\sum\limits_{n=1}^{d}
\frac{1}{m} \sum\limits_{i=1}^{m}\Big[\textrm{Tr}(P_{i,n}^{(b)}|\psi_{i}\rangle\langle\psi_{i}|)\Big]^2\\
&~~~~~~=\frac{1}{m}\sum\limits_{i=1}^{m}\sum\limits_{b=1}^{M}\sum\limits_{n=1}^{d}\Big[\textrm{Tr}(P_{i,n}^{(b)}|\psi_{i}\rangle\langle\psi_{i}|)\Big]^2.
\end{array}
\end{equation}
By using the relation (\ref{M-1})
for pure state $\rho$, we obtain
\begin{equation}\label{}
J(\rho)\leq \frac{M-1}{d}+ \frac{1}{m}\sum\limits_{i=1}^{m}\kappa_{i}.
\end{equation}
The inequality holds for mixed states since $J(\rho)$ is a linear function. This completes the proof. ~~~~~~~~~~~~~~~~~~~~~~ $\square$

Especially, when we use the complete sets of MUMs, that is, $M=d+1$, the inequality  becomes
\begin{equation}
J(\rho)\leq 1+ \frac{1}{m}\sum\limits_{i=1}^{m}\kappa_{i}.
\end{equation}
What's more, when the efficiencies of each set of MUMs are same, the right-hand side of the inequality becomes $1+\kappa$, and the criterion in Ref. \cite{PRA.89.064302} is the special case of our criterion when $m=2$.  When $m=2$ and $\kappa=1$, our criterion (of Theorem 1) reduces to the previous one in Ref. \cite{PRA.86.022311}, which demonstrates that $J(\rho)\leq 2$ for all separable states $\rho$ in $\mathbb{C}^d \otimes \mathbb{C}^{d}$, if there exists a complete set of MUBs in $\mathbb{C}^d$.

 For two qudit systems, the criterion in Ref. \cite{PRA.86.022311} is shown to be powerful in detecting entanglement of particular states, but  when $d$ is not a prime power, the criterion in Ref. \cite{PRA.86.022311} becomes less effective, since the existence of a complete set of MUBs remains open for Hilbert spaces of nonprime power dimension. The authors of Ref. \cite{PRA.89.064302} showed that their criterion is more efficient than the criterion in Ref. \cite{PRA.86.022311} and detects all the entangled isotropic states of arbitrary dimension $d$. As the special case of our criterion when $m=2$, the criterion in Ref. \cite{PRA.89.064302} can only be used to $d$-dimensional bipartite systems and two sets of $d+1$ MUMs on $\mathbb{C}^d$ with the same parameter $\kappa$, while our criterion of Theorem 1 can be used to arbitrary $d$-dimensional $m$-partite systems ($m\geq 2$) and $m$ sets of $M$ MUMs on $\mathbb{C}^{d}$ with different efficiencies $\kappa_{i}$, thus our criterion is of the advantages of more effective and wider application range.

For the bipartite system and multipartite system of subsystems with different dimensions, we have no idea how to detect the separability of  states  using complete sets of MUMs, but with incomplete sets of MUMs, we have the following conclusions.

\vspace{0.2cm} \noindent\textbf{Theorem 2.}~~{\slshape Let $\rho$ be a density matrix in $\mathbb{C}^{d_{1}} \otimes \mathbb{C}^{d_{2}}$, and $\{\mathcal{P}^{(b)}\}_{b=1}^{M}$ and
$\{\mathcal{Q}^{(b)}\}_{b=1}^{M}$ be any two sets of $M$ $MUMs$ on $\mathbb{C}^{d_{1}}$ and $\mathbb{C}^{d_{2}}$ with efficiency $\kappa_{1}$, $\kappa_{2}$, respectively, where $\mathcal{P}^{(b)}=\{P_{n}^{(b)}\}_{n=1}^{d_{1}}$, and $\mathcal{Q}^{(b)}=\{Q_{n'}^{(b)}\}_{n'=1}^{d_{2}}$, $b=1,2,...,M$.
Define
\begin{equation}
\begin{array}{ll}
J(\rho)=\max\limits_{\begin{subarray}{c} \{P_{n_i}^{(b)}\}_{i=1}^{d}\subseteq\mathcal{P}^{(b)} \\ \{Q_{n_i}^{(b)}\}_{i=1}^{d}\subseteq\mathcal{Q}^{(b)} \\ 1\leq b\leq M \end{subarray}}
\sum\limits_{b=1}^{M}\sum\limits_{i=1}^{d}
\mathrm{Tr}(P_{n_i}^{(b)}\otimes Q_{n_i}^{(b)}\rho).
\end{array}
\end{equation}
Here $d=\min\{d_{1},d_{2}\}$.
If $\rho$ is separable, then
\begin{equation}
J(\rho)\leq\frac{1}{2}\Big[(M-1)\Big(\frac{1}{d_{1}}+\frac{1}{d_{2}}\Big)+\kappa_{1}+\kappa_{2}\Big].
\end{equation}
}
\vspace{0.2cm}

\noindent\textbf{Proof.}~We need only consider a pure separable state $\rho=|\phi\rangle\langle\phi|\otimes|\psi\rangle\langle\psi|$,
since $\sum\limits_{b=1}^{M}\sum\limits_{i=1}^{d}
\textrm{Tr}(P_{n_i}^{(b)}\otimes Q_{n_i}^{(b)}\rho)$ is a linear function of $\rho$. We have
\begin{equation}
\begin{array}{ll}
&~~~~~\sum\limits_{b=1}^{M}\sum\limits_{i=1}^{d}
               \textrm{Tr}(P_{n_i}^{(b)}\otimes Q_{n_i}^{(b)}\rho)\\
&=\sum\limits_{b=1}^{M}\sum\limits_{i=1}^{d}
   \textrm{Tr}(P_{n_i}^{(b)}|\phi\rangle\langle\phi|) \textrm{Tr}(Q_{n_i}^{(b)}|\psi\rangle\langle\psi|)\\
&\leq \sum\limits_{b=1}^{M}\sum\limits_{i=1}^{d}
      \frac{1}{2}\{[\textrm{Tr}(P_{n_i}^{(b)}|\phi\rangle\langle\phi|)]^{2}+[\textrm{Tr}(Q_{n_i}^{(b)}|\psi\rangle\langle\psi|)]^{2}\}\\
&=\frac{1}{2}\sum\limits_{b=1}^{M}\sum\limits_{i=1}^{d}[\textrm{Tr}(P_{n_i}^{(b)}|\phi\rangle\langle\phi|)]^{2}
+\frac{1}{2}\sum\limits_{b=1}^{M}\sum\limits_{i=1}^{d}[\textrm{Tr}(Q_{n_i}^{(b)}|\psi\rangle\langle\psi|)]^{2}\\
&\leq\frac{1}{2}[\frac{M-1}{d_{1}}+\frac{1-\kappa_{1}+(\kappa_{1}d_{1}-1)\textrm{Tr}(|\phi\rangle\langle\phi|)^{2}}{d_{1}-1}
+\frac{M-1}{d_{2}}+\frac{1-\kappa_{2}+(\kappa_{2}d_{2}-1) \textrm{Tr}(|\psi\rangle\langle\psi|)^{2}}{d_{2}-1}]\\
&=\frac{1}{2}[(M-1)(\frac{1}{d_{1}}+\frac{1}{d_{2}})+\kappa_{1}+\kappa_{2}],
\end{array}
\end{equation}
where the inequality (\ref{M-1}) is used. This completes the proof.  ~~~~~~~~~~~~~~~~~~~~~~~~~~~~~~~~~~~~~~~~~~~~~~~~~~~~~~~~~~~~~~~~~~~~~~~~~~~~~ $\square$

It is worthy to note that the criterion in Ref. \cite{PRA.89.064302} is the corollary of Theorem 2. In fact, if  $d_1=d_2=d$, and $\{\mathcal{P}^{(b)}\}_{b=1}^{M}$ and
$\{\mathcal{Q}^{(b)}\}_{b=1}^{M}$ are any two sets of $d+1$ MUMs on $\mathbb{C}^{d}$ with the same efficiency $\kappa$, then by Theorem 2 there is
\begin{equation}\label{}
J(\rho)=\sum\limits_{b=1}^{d+1}\sum\limits_{i=1}^{d}
\textrm{Tr}(P_{i}^{(b)}\otimes Q_{i}^{(b)}\rho)\leq 1+\kappa,
\end{equation}
which is the desired result. Therefore, the criterion in Ref. \cite{PRA.89.064302} is the special case of our criterion of Theorem 2.

Just as noted in Ref. \cite{PRA.89.064302}, the entanglement detection based on MUMs is more efficient than the one based on MUBs  for some states.  Our criteria (Theorems 1 and 2) and the criterion in Ref. \cite{PRA.89.064302} as the special case of Theorems 1 and 2,  are both necessary and sufficient for the separability of the isotropic states, namely, they can detect all the entanglement of the isotropic states. It should be emphasized that, unlike the criterion based on MUBs in Ref. \cite{PRA.86.022311}, our criteria work perfectly for any dimension $d$.

By using the Cauchy-Schwarz inequality, we can obtain stronger bound  than  that in Theorem 2.

\vspace{0.2cm} \noindent\textbf{Theorem 3.}~~{\slshape Let $\rho$ be a density matrix in $\mathbb{C}^{d_{1}} \otimes \mathbb{C}^{d_{2}}$, and $\{\mathcal{P}^{(b)}\}_{b=1}^{M}$ and
$\{\mathcal{Q}^{(b)}\}_{b=1}^{M}$ be any two sets of $M$ $MUMs$ on $\mathbb{C}^{d_{1}}$ and $\mathbb{C}^{d_{2}}$ with efficiency $\kappa_{1}$, $\kappa_{2}$, respectively, where $\mathcal{P}^{(b)}=\{P_{n}^{(b)}\}_{n=1}^{d_{1}}$, and $\mathcal{Q}^{(b)}=\{Q_{n'}^{(b)}\}_{n'=1}^{d_{2}}$, $b=1,2,...,M$.
Define
\begin{equation}
\begin{array}{ll}
J(\rho)=\max\limits_{\begin{subarray}{c} \{P_{n_i}^{(b)}\}_{i=1}^{d}\subseteq\mathcal{P}^{(b)} \\ \{Q_{n_i}^{(b)}\}_{i=1}^{d}\subseteq\mathcal{Q}^{(b)} \\ 1\leq b\leq M \end{subarray}}
\sum\limits_{b=1}^{M}\sum\limits_{i=1}^{d}
\mathrm{Tr}(P_{n_i}^{(b)}\otimes Q_{n_i}^{(b)}\rho).
\end{array}
\end{equation}
Here $d=\min\{d_{1},d_{2}\}$.
If $\rho$ is separable, then
\begin{equation}
J(\rho)\leq\sqrt{\frac{M-1}{d_{1}}+\kappa_{1}}\sqrt{\frac{M-1}{d_{2}}+\kappa_{2}}.
\end{equation}
}
\vspace{0.2cm}

\noindent\textbf{Proof.}~We need only consider a pure separable state $\rho=|\phi\rangle\langle\phi|\otimes|\psi\rangle\langle\psi|$,
since $\sum\limits_{b=1}^{M}\sum\limits_{i=1}^{d}
\textrm{Tr}(P_{n_i}^{(b)}\otimes Q_{n_i}^{(b)}\rho)$ is a linear function of $\rho$. We have
\begin{equation}
\begin{array}{ll}
&~~~~~\sum\limits_{b=1}^{M}\sum\limits_{i=1}^{d}
               \textrm{Tr}(P_{n_i}^{(b)}\otimes Q_{n_i}^{(b)}\rho)\\
&=\sum\limits_{b=1}^{M}\sum\limits_{i=1}^{d}
   \textrm{Tr}(P_{n_i}^{(b)}|\phi\rangle\langle\phi|) \textrm{Tr}(Q_{n_i}^{(b)}|\psi\rangle\langle\psi|)\\
&\leq \sqrt{\sum\limits_{b=1}^{M}\sum\limits_{i=1}^{d_{1}}[\textrm{Tr}(P_{n_i}^{(b)}|\phi\rangle\langle\phi|)]^{2}}
      \sqrt{\sum\limits_{b=1}^{M}\sum\limits_{i=1}^{d_{2}}[\textrm{Tr}(Q_{n_i}^{(b)}|\psi\rangle\langle\psi|))]^{2}}\\
&\leq\sqrt{\frac{M-1}{d_{1}}+\frac{1-\kappa_{1}+(\kappa_{1}d_{1}-1)\textrm{Tr}(|\phi\rangle\langle\phi|)^{2}}{d_{1}-1}}
     \sqrt{\frac{M-1}{d_{2}}+\frac{1-\kappa_{2}+(\kappa_{2}d_{2}-1) \textrm{Tr}(|\psi\rangle\langle\psi|)^{2}}{d_{2}-1}}\\
&=\sqrt{\frac{M-1}{d_{1}}+\kappa_{1}}\sqrt{\frac{M-1}{d_{2}}+\kappa_{2}},
\end{array}
\end{equation}
where the Cauchy-Schwarz inequality and the inequality (\ref{M-1}) are used.
This completes the proof.  ~~~~~~~~~~~~~~~~~~~~~~~~~~ $\square$

The bound in Theorem 3 is lower than that in Theorem 2 since $\sqrt{\frac{M-1}{d_{1}}+\kappa_{1}}\sqrt{\frac{M-1}{d_{2}}+\kappa_{2}}\leq \frac{1}{2}\big(\frac{M-1}{d_{1}}+\frac{M-1}{d_{2}}+\kappa_{1}+\kappa_{2}\big)$.

The Proposition 6 in Ref.\cite{1407.7333} is the special case $d_1=d_2=d$ of Theorem 3. It detects all the entanglement of isotropic states for arbitrary dimension $d$, so does Theorem 3.

\vspace{0.2cm} \noindent\textbf{Theorem 4.}~~{\slshape
Suppose that $\rho$ is a density matrix in $\mathbb{C}^{d_{1}}\otimes\mathbb{C}^{d_{2}}\otimes\cdots\otimes\mathbb{C}^{d_{m}} $ and $\mathcal{P}^{(b)}_{i}$ are any sets of $M$ MUMs on $\mathbb{C}^{d_{i}}$ with the efficiencies $\kappa_{i}$,
where  $b=1,2,\cdots,M$, $i=1,2,\cdots,m$.
Let $d=\min \{d_{1},d_{2},\cdots,d_{m}\},$ and
define
\begin{equation}\label{}
J(\rho)=\max_{\begin{subarray}{c} \{P_{i,n}^{(b)}\}_{n=1}^{d}\subseteq\mathcal{P}^{(b)}_{i} \\ i=1,2,\cdots,m \\ b=1,2,\cdots,M \end{subarray}}
\sum\limits_{b=1}^{M}\sum\limits_{n=1}^{d}
\mathrm{Tr}\Big(\big(\otimes^{m}_{i=1} P_{i,n}^{(b)}\big) \rho\Big).
\end{equation}
If $\rho$ is fully separable, then
\begin{equation} \label{M-3}
J(\rho)\leq \frac{1}{m}\sum\limits_{i=1}^{m}\Big(\frac{M-1}{d_{i}}+\kappa_{i}\Big).
\end{equation}
}

\vspace{0.2cm}

\noindent\textbf{Proof.}~Let $\rho=\sum\limits_{j}p_{j}\rho_{j}$ with $\sum\limits_{j}p_{j}=1$, be a fully separable density matrix, where $\rho_{j}=\otimes_{i=1}^{m}\rho_{ij}$.
Since
\begin{equation}
\begin{array}{ll}
&\sum\limits_{b=1}^{M}\sum\limits_{n=1}^{d}
              \textrm{Tr}[(\otimes^{m}_{i=1} P_{i,n}^{(b)})\rho_j]\\
=&\sum\limits_{b=1}^{M}\sum\limits_{n=1}^{d}
              \textrm{Tr}[\otimes^{m}_{i=1} (P_{i,n}^{(b)}\rho_{ij})]\\
=&\sum\limits_{b=1}^{M}\sum\limits_{n=1}^{d}
              \prod_{i=1}^{m}\textrm{Tr}(P_{i,n}^{(b)}\rho_{ij})\\
\leq&\sum\limits_{b=1}^{M}\sum\limits_{n=1}^{d}
              [\frac{1}{m}\sum\limits_{i=1}^{m}\big(\textrm{Tr}(P_{i,n}^{(b)}\rho_{ij})\big)^{2}]^{\frac{m}{2}}\\
\leq&\sum\limits_{b=1}^{M}\sum\limits_{n=1}^{d}
              [\frac{1}{m}\sum\limits_{i=1}^{m}\big(\textrm{Tr}(P_{i,n}^{(b)}\rho_{ij})\big)^{2}]\\
\leq&\frac{1}{m}\sum\limits_{i=1}^{m}\sum\limits_{b=1}^{M}\sum\limits_{n_{i}=1}^{d_{i}}\big(\textrm{Tr}(P_{i,n_{i}}^{(b)}\rho_{ij})\big)^{2}\\
\leq&\frac{1}{m}\sum\limits_{i=1}^{m}(\frac{M-1}{d_{i}}+\kappa_{i}),
\end{array}
\end{equation}
there is
\begin{equation}
\begin{array}{ll}
&\sum\limits_{b=1}^{M}\sum\limits_{n=1}^{d}
              \textrm{Tr}[(\otimes^{m}_{i=1} P_{i,n}^{(b)})\rho]\\
=&\sum\limits_{j}p_j \sum\limits_{b=1}^{M}\sum\limits_{n=1}^{d}
              \textrm{Tr}[(\otimes^{m}_{i=1} P_{i,n}^{(b)})\rho_j]\\
\leq&\frac{1}{m}\sum\limits_{i=1}^{m}(\frac{M-1}{d_{i}}+\kappa_{i}),
\end{array}
\end{equation}
which implies that inequality (\ref{M-3}) holds. It is complete.   ~~~~~~~~~~~~~~~~~~~~~~~~~~~~~~~~~~~~~~~~~~~~~~~~~~~~~~~~~~~~~~~~~~~~~~~~~~ $\square$

\vspace{0.2cm} \noindent\textbf{Theorem 5.}~~{\slshape
Suppose that $\rho$ is a density matrix in $\mathbb{C}^{d_{1}}\otimes\mathbb{C}^{d_{2}}\otimes\cdots\otimes\mathbb{C}^{d_{m}} $ and $\mathcal{P}^{(b)}_{i}$ are any sets of $M$ MUMs on $\mathbb{C}^{d_{i}}$ with the efficiencies $\kappa_{i}$,
where  $b=1,2,\cdots,M$, $i=1,2,\cdots,m$.
Let $d=\min \{d_{1},d_{2},\cdots,d_{m}\},$ and
define
\begin{equation}\label{}
J(\rho)=\max_{\begin{subarray}{c} \{P_{i,n}^{(b)}\}_{n=1}^{d}\subseteq\mathcal{P}^{(b)}_{i} \\ i=1,2,\cdots,m \\ b=1,2,\cdots,M \end{subarray}}
\sum\limits_{b=1}^{M}\sum\limits_{n=1}^{d}
\mathrm{Tr}\Big(\big(\otimes^{m}_{i=1} P_{i,n}^{(b)}\big) \rho\Big).
\end{equation}
If $\rho$ is fully separable, then
\begin{equation} \label{M-3}
J(\rho)\leq \min_{1\leq i\neq j\leq m}\sqrt{\frac{M-1}{d_{i}}+\kappa_{i}}\sqrt{\frac{M-1}{d_{j}}+\kappa_{j}}.
\end{equation}
}

\vspace{0.2cm}

\noindent\textbf{Proof.}~Let $\rho=\sum\limits_{k}p_{k}\rho_{k}$ with $\sum\limits_{k}p_{k}=1$, be a fully separable density matrix, where $\rho_{k}=\otimes_{i=1}^{m}\rho_{ik}$. For any $i,j\in\{1, 2, \cdots, m \}$ and $i\neq j$,
Since
\begin{equation}
\begin{array}{ll}
&\sum\limits_{b=1}^{M}\sum\limits_{n=1}^{d}
              \textrm{Tr}[(\otimes^{m}_{i=1} P_{i,n}^{(b)})\rho_k]\\
=&\sum\limits_{b=1}^{M}\sum\limits_{n=1}^{d}
              \prod_{i=1}^{m}\textrm{Tr}(P_{i,n}^{(b)}\rho_{ik})\\
\leq&\sum\limits_{b=1}^{M}\sum\limits_{n=1}^{d}\textrm{Tr}(P_{i,n}^{(b)}\rho_{ik})\textrm{Tr}(P_{j,n}^{(b)}\rho_{jk})\\
\leq&\sqrt{\sum\limits_{b=1}^{M}\sum\limits_{n=1}^{d_{i}}(\textrm{Tr}(P_{i,n}^{(b)}\rho_{ik}))^{2}}
     \sqrt{\sum\limits_{b=1}^{M}\sum\limits_{n=1}^{d_{j}}(\textrm{Tr}(P_{j,n}^{(b)}\rho_{jk}))^{2}}\\
\leq&\sqrt{\frac{M-1}{d_{i}}+\kappa_{i}}\sqrt{\frac{M-1}{d_{j}}+\kappa_{j}},
\end{array}
\end{equation}
where we have used the Cauchy-Schwarz inequality and the relation (\ref{M-1}),
there is
\begin{equation}
\begin{array}{ll}
&\sum\limits_{b=1}^{M}\sum\limits_{n=1}^{d}
              \textrm{Tr}[(\otimes^{m}_{i=1} P_{i,n}^{(b)})\rho]\\
=&\sum\limits_{k}p_k \sum\limits_{b=1}^{M}\sum\limits_{n=1}^{d}
              \textrm{Tr}[(\otimes^{m}_{i=1} P_{i,n}^{(b)})\rho_k]\\
\leq&\sqrt{\frac{M-1}{d_{i}}+\kappa_{i}}\sqrt{\frac{M-1}{d_{j}}+\kappa_{j}},
\end{array}
\end{equation}
which implies that inequality (\ref{M-3}) holds. It is complete.   ~~~~~~~~~~~~~~~~~~~~~~~~~~~~~~~~~~~~~~~~~~~~~~~~~~~~~~~~~~~~~~~~~~~~~~~~~~ $\square$

For Theorems 4 and 5, we don't require the subsystems with the same dimension, so we can use them straightforward to detect $k$-nonseparable states with respect to a fixed partition. For an $N$-partite state $\rho$ in $\mathbb{C}_1\otimes\mathbb{C}_2\otimes\cdots\otimes\mathbb{C}_N=\mathbb{C}^{d_{1}}\otimes\mathbb{C}^{d_{2}}\otimes\cdots\otimes\mathbb{C}^{d_{k}}$, if there are sets of  $M$ MUMs $\{\mathcal{P}^{(b)}_{i}\}_{b=1}^M$  on $\mathbb{C}^{d_{i}}$ with the efficiencies $\kappa_{i}$ such that $\sum\limits_{b=1}^{M}\sum\limits_{n=1}^{d}
\textrm{Tr}\Big(\big(\bigotimes^{k}_{i=1} P_{i,n}^{(b)}\big) \rho\Big)> \frac{1}{k}\sum_{i=1}^{k}\Big(\frac{M-1}{d_{i}}+\kappa_{i}\Big)$, or $\sum\limits_{b=1}^{M}\sum\limits_{n=1}^{d}
\textrm{Tr}\Big(\big(\bigotimes^{k}_{i=1} P_{i,n}^{(b)}\big) \rho\Big) >\min\limits_{1\leq i\neq j\leq m}\sqrt{\frac{M-1}{d_{i}}+\kappa_{i}}\sqrt{\frac{M-1}{d_{j}}+\kappa_{j}}$ for some $\{P_{i,n}^{(b)}\}_{n=1}^{d}\subseteq\mathcal{P}^{(b)}_{i}$, then $\rho$ is $k$-nonseparable in $\mathbb{C}^{d_{1}}\otimes\mathbb{C}^{d_{2}}\otimes\cdots\otimes\mathbb{C}^{d_{k}}$, that is, $\rho$ can not be written as a  convex combination of $N$-partite pure state each of which is $k$-separable in $\mathbb{C}^{d_{1}}\otimes\mathbb{C}^{d_{2}}\otimes\cdots\otimes\mathbb{C}^{d_{k}}$, where  $d=\min \{d_{1},d_{2},\cdots,d_{k}\}$, and $i=1,2,\cdots,k$.

Our criteria are much better than the previous ones  in Ref.\cite{PRA.86.022311, PRA.89.064302, 1407.7333}.  First, the criterion in Ref.\cite{PRA.89.064302}, the Propositions 2 and 6 in Ref.\cite{1407.7333}, and inequality (8) in Ref. \cite{PRA.86.022311} are the special cases of our criteria for two-qudit systems. Second,  the authors of Ref.\cite{PRA.89.064302, 1407.7333}
only  provided separability criteria for a bipartite system of two $d$-dimensional subsystems, while we present separability criteria to detect entanglement of quantum states in $(\mathbb{C}^{d})^{\otimes m}$, $\mathbb{C}^{d_{1}} \otimes \mathbb{C}^{d_{2}}$, and  $\mathbb{C}^{d_{1}}\otimes\mathbb{C}^{d_{2}}\otimes\cdots\otimes\mathbb{C}^{d_{m}} $,
where  $m\geq 2$, that is, the criteria in Ref.\cite{PRA.89.064302, 1407.7333} are applied to bipartite systems of two subsystems with same dimension, while our separability can be used to not only bipartite systems of two subsystems with same dimension but also multipartite qudit systems  and multipartite systems of subsystems with different dimensions. Third, unlike the criterion Ref.\cite{PRA.86.022311} based on MUBs,  our criteria and the criteria in Ref.\cite{PRA.89.064302, 1407.7333}  detect all the entangled isotropic states of
arbitrary dimension $d$. The powerfulness of the criteria based on MUMs is due to the fact that there always exists a complete set of MUMs, which is not the case for MUBs when $d$ is not a prime power.  Last, our criteria can be applied to detect $k$-nonseparability of $N$-partite systems ($N>2$, $2<k\leq N$), while the criteria in Ref.\cite{PRA.86.022311, PRA.89.064302, 1407.7333} can not.

\vspace{0.5cm} \noindent{\bf\large Conclusion and discussions }

\noindent
In summary we have investigated the entanglement detection using mutually unbiased measurements and presented separability criteria for
multipartite systems composed of $m$ $d$-dimensional subsystems,  bipartite systems composed of a $d_1$-dimensional subsystem and a $d_2$-dimensional subsystem, and  multipartite systems of $m$ multi-level subsystems via mutually unbiased measurements, where $m\geq 2$. These criteria are of the advantages of more effective and wider application range than previous criteria.  They  provide experimental implementation in detecting entanglement of unknown quantum states, and  are beneficial for experiments since they require only a few local measurements. One can flexibly use them in practice. For multipartite systems, the definition of separability is not unique. We can detect the $k$-nonseparability of  $N$-partite and high dimensional systems.
It would be interesting to study the separability criterion of multipartite systems with different dimensions via complete set of MUMs.

\vspace{0.5cm} \noindent{\bf\large Acknowledgments }

\noindent
This work was supported by the National Natural Science Foundation
of China under Grant Nos: 11371005, 11475054; the Hebei Natural Science Foundation
of China under Grant Nos: A2012205013, A2014205060.

\vspace{0.5cm} \noindent{\bf\large Author contributions }

\noindent
L.L., T.G. and F.Y. contributed equally to this work.    All authors wrote the main manuscript text and reviewed the manuscript.

\vspace{0.5cm} \noindent{\bf\large Additional information }

\noindent
Competing financial interests: The authors declare no competing financial interests.

\begin{thebibliography}{99}
\bibitem{PRL70.1895} Bennett, C. H. \textit{et al}. Teleporting an unknown quantum state via dual classical and Einstein-Podolsky-Rosen channels. \textit{Phys. Rev. Lett.} \textbf{70}, 1895 (1993).
\bibitem{PRL67.661} Ekert, A. K. Quantum cryptography based on Bell’s theorem. \textit{Phys. Rev. Lett.} \textbf{67}, 661 (1991).
\bibitem{PRL68.557} Bennett, C. H., Brassard, G. $\&$ Mermin, N. D. Quantum cryptography without Bell’s theorem. \textit{Phys. Rev. Lett.} \textbf{68}, 557 (1992).
\bibitem{PR448.1} Wang, X. B., Hiroshima, T., Tomita, A. $\&$ Hayashi, M. Quantum information with Gaussian states. \textit{Phys. Rep.} \textbf{448}, 1 (2007).
\bibitem{Science283.2050} Lo, H. K. $\&$ Chau, H. F. Unconditional security of quantum key distribution over arbitrarily long distances. \textit{Science} \textbf{283}, 2050 (1999).
\bibitem{PJB41.75} Yan, F. L. $\&$ Zhang, X. Q. A scheme for secure direct communication using EPR pairs and teleportation. \textit{Eur. Phys. J. B} \textbf{41}, 75 (2004).
\bibitem{JPA38.5761} Gao, T., Yan, F. L. $\&$ Wang, Z. X. Deterministic secure direct communication using GHZ states and swapping quantum entanglement. \textit{J. Phys. A} \textbf{38}, 5761 (2005).
\bibitem{PRA83.022319} Yan, F. L., Gao, T. $\&$ Chitambar, E. Two local observables are sufficient to characterize maximally entangled states of $N$ qubits. \textit{Phys. Rev. A} \textbf{83}, 022319 (2011).
\bibitem{GaoEPL84} Gao, T., Yan, F. L. $\&$ Li, Y. C. Optimal controlled teleportation. \textit{Europhys. Lett.} \textbf{84}, 50001 (2008).
\bibitem{Nature404.247} Bennett, C. H. $\&$ DiVincenzo, D. P. Quantum information and computation. \textit{Nature} \textbf{404}, 247 (2000).
\bibitem{77.2816} Deutsch, D. \textit{et al}. Quantum privacy amplification and the security of quantum cryptography over noisy channels. \textit{Phys. Rev. Lett.} \textbf{77}, 2818 (1996).
\bibitem{56.1163} Fuchs, C. A., Gisin, N., Griffiths, R. B., Niu, C. S. $\&$ Peres, A. Optimal eavesdropping in quantum cryptography. I. Information bound and optimal strategy. \textit{Phys. Rev. A} \textbf{56}, 1163 (1997).
\bibitem{PLA276.8} Albeberio, S. $\&$ Fei, S. M. Teleportation of general finite-dimensional quantum systems. \textit{Phys. Lett. A} \textbf{276}, 8 (2000).
\bibitem{272.32} D'Ariano, G. M., Presti, P. L. $\&$ Sacchi, M. F. Bell measurements and observables. \textit{Phys. Lett. A} \textbf{272}, 32 (2000).
\bibitem{PRA66.012301} Albeverio, S., Fei, S. M. $\&$ Yang, W. L. Optimal teleportation based on Bell measurements. \textit{Phys. Rev. A} \textbf{66}, 012301 (2002).
\bibitem{GaoCTP2004} Gao, T. Quantum logic networks for probabilistic and controlled teleportation of unknown quantum states. \textit{Commun. Theor. Phys.} \textbf{42}, 223 (2004).
\bibitem{PRL69.2881} Bennett, C. H. $\&$ Wiesner, S. J. Communication via one- and two-particle operators on Einstein-Podolsky-Rosen states. \textit{Phys. Rev. Lett.} \textbf{69}, 2881 (1992).
\bibitem{PRL77.1413} Peres, A. Separability criterion for density matrices. \textit{Phys. Rev. Lett.} \textbf{77}, 1413 (1996).
\bibitem{QIC3.193} Chen, K. $\&$ Wu, L. A. A matrix realignment method for recognizing entanglement. \textit{Quantum Inf. Comput.} \textbf{3}, 193 (2003).
\bibitem{PRA59.4206} Horodecki, M. $\&$ Horodecki, P. Reduction criterion of separability and limits for a class of distillation protocols. \textit{Phys. Rev. A.} \textbf{59}, 4206 (1999).
\bibitem{PRL99.130504} G\"{u}hne, O., Hyllus, P., Gittsovich, O. $\&$ Eisert, J. Covariance matrices and the separability problem. \textit{Phys. Rev. Lett.} \textbf{99}, 130504 (2007).
\bibitem{PRA82.062113} Gao, T. $\&$ Hong, Y. Detection of genuinely entangled and nonseparable $n$-partite quantum states. \textit{Phys. Rev. A} \textbf{82}, 062113 (2010).
\bibitem{EPL.104.20007} Gao, T., Hong, Y., Lu, Y. $\&$ Yan, F. L. Efficient $k$-separability criteria for mixed multipartite quantum states. \textit{Europhys. Lett.} \textbf{104}, 20007 (2013).
\bibitem{PRA.86.062323} Hong, Y., Gao, T. $\&$ Yan, F. L. Measure of multipartite entanglement with computable lower bounds. \textit{Phys. Rev. A } \textbf{86}, 062323 (2012).
\bibitem{PRL.112.180501} Gao, T., Yan, F. L. $\&$ van Enk, S. J. Permutationally invariant part of a density matrix and nonseparability of $N$-qubit states. \textit{Phys. Rev. Lett.} \textbf{112}, 180501 (2014).
\bibitem{1409.3386[7]} Schwinger, J. Unitary operator bases. \textit{Pro. Nat. Acad. Sci. U. S. A.} \textbf{46}, 570-579 (1960).
\bibitem{1409.3386[11]} Wootters, W. K. $\&$ Fields, B. D. Optimal state-determination by mutually unbiased measurements. \textit{Ann. Phys.} (N. Y.) \textbf{191}, 363 (1989).
\bibitem{PRL.110.143601[3]} Wehner, S. $\&$ Winter, A. Entropic uncertainty relations-a survey. \textit{New J. Phys.} \textbf{12}, 025009 (2010).
\bibitem{PRL.110.143601[4]} Barnett, S. M. Quantum Information (Oxford University Press, Oxford, England, 2009).
\bibitem{PRL.110.143601[5]} Durt, T., Englert, B. G., Bengtsson, I. $\&$ \.{Z}yczkowski, K. On mutually unbiased bases. \textit{Int. J. Quantum. Inform.} \textbf{08}, 535 (2010).
\bibitem{NP.6.659} Berta, M. \textit{et al}. The uncertainty principle in the presence of quantum memory. \textit{Nature Physics} \textbf{6}, 659 (2010).
\bibitem{PRA.90.062127} Berta, M., Coles, P. J. $\&$ Wehner, S. Entanglement-assisted guessing of complementary measurement outcomes. \textit{Phys. Rev. A} \textbf{90}, 062127 (2014).
\bibitem{NP.7.757} Prevedel, R. \textit{et al}. Experimental investigation of the uncertainty principle in the presence of quantum memory and its application to witnessing entanglement. \textit{Nature Physics} \textbf{7}, 757 (2011).
\bibitem{PRA.86.022311} Spengler, C., Huber, M., Brierley, S., Adaktylos, T. $\&$ Hiesmayr, B. C. Entanglement detection via mutually unbiased bases. \textit{Phys. Rev. A} \textbf{86}, 022311 (2012).
\bibitem{NJP.16.053038} Kalev, A. $\&$ Gour, G. Mutually unbiased measurements in finite dimensions. \textit{New J. Phys.} \textbf{16}, 053038 (2014).
\bibitem{PRA.89.064302} Chen, B., Ma, T. $\&$ Fei, S. M. Entanglement detection using mutually unbiased measurements. \textit{Phys. Rev. A} \textbf{89}, 064302 (2014).
\bibitem{1407.7333} Rastegin, A. E. On uncertainty relations and entanglement detection with mutually unbiased measurements. \textit{Open Sys. $\&$ Inf. Dyn.} \textbf{22}, 1550005 (2015).
\bibitem{1407.6816} Chen, B. $\&$ Fei, S. M. Uncertainty relations based on mutually unbiased measurements. \textit{e-print} arXiv: 1407.6816 [quant-ph] (2014).
\bibitem{website} Hardy, G., Littlewood, J. E. $\&$ P\'{o}lya, G. \textit{Inequalties} (Cambridge University Press, Cambridge, UK, 1952)

\end{thebibliography}
\end{document}